\documentclass{article}
\usepackage{graphicx}
\usepackage{balance}
\usepackage{url}
\usepackage{amssymb}
\pdfoutput=1

\newcommand{\ttilde}{{\raise0.5ex\hbox{$\scriptstyle\sim$}}}

\begin{document}

\title{Hopping over Big Data: Accelerating Ad-hoc OLAP Queries with Grasshopper Algorithms}

\author{
Alexander Russakovsky
       \thanks{alex.russakovsky@huawei.com,Huawei US R\&D Center, Santa Clara, USA}
}

\maketitle

\begin{abstract}
This paper presents a family of algorithms for fast subset filtering within 
ordered sets of integers representing composite keys. Applications include 
significant acceleration of (ad-hoc) analytic queries against a data warehouse 
without any additional indexing. The algorithms work for point, range and set 
restrictions on multiple attributes, in any combination, and are inherently 
multidimensional. The main idea consists in intelligent combination of 
sequential crawling with jumps over large portions of irrelevant keys. 
The way to combine them is adaptive to characteristics of the underlying 
data store.
\end{abstract}

\section{Introduction}

Business Intelligence (BI) applications, in their quest to provide decision support, 
rely on OLAP and Data Mining techniques to extract actionable information from 
collected data.
Modern applications are characterized by massive volumes of data
and extensive empowerment of users resulting in requirements to ensure 
fast response for ad-hoc queries against this data. 

Classical relational data warehousing techniques \cite{kimball:book96}
are often combined or replaced nowadays with non-relational distributed processing 
systems. One way to interpret this trend in the context of ad-hoc queries is to observe 
that maintaining adequate indexing for massive data volumes becomes impractical and
perhaps the only remaining strategy for ad-hoc queries is the ``brute force" approach, 
i.e. full scan of data distributed over a large number of nodes. Performance improvements
can be achieved this way, but at a high price.

Scalability and performance requirements have always been critical for BI. Multidimensional
OLAP techniques have been used to address performance problems, but scalability had been limited. 
Other popular ways to address the ad-hoc query performance problems have been in-memory
and columnar databases. All of these techniques are beneficial and some work better than the others
for ad-hoc OLAP queries.

The family of algorithms suggested in this paper, ``Grasshopper algorithms",  is aimed at acceleration 
of ad-hoc OLAP queries without any additional indexing. They work for point, range and set restrictions
on dimensional attributes, in any combination. The algorithms combine sequential scanning 
with long hops over irrelevant data; in the majority of cases they perform much faster than full scan, 
but practically never worse. A grasshopper can crawl or can jump quite far. This explains the name.

An {\it OLAP cube} is a unit of logical data organization reflecting a vector functional dependency $F$ in 
which independent variables are called dimensions, and dependent variables are called measures. Dimensions often
have hierarchical structure induced by additional (scalar) functional dependencies on independent variables. Various aspects
of OLAP have been extensively studied in the literature (see e.g. \cite{chaudhuri:overview, kimball:book96, pedersen:overview} 
and references in there, as well as on-line OLAP bibliography \cite{olap:bibliography}).

An {\it ad-hoc OLAP query} is a query against the cube in which various filters may be placed on some of the participating 
variables, and measure values may have to be aggregated. Since it is not known in advance, to which variables 
filters will be applied, and what restrictions the filter will pose, techniques like materialized views may
not be helpful.

OLAP implementations tend to use dictionaries to encode dimensional attribute values with surrogate keys. We will
assume that all surrogate keys are integers. For unordered attributes we prefer these integers to be consecutive;
ordered attributes with integer values do not have to be encoded, but if any naturally ordered attribute is encoded,
encoding must preserve the order. We then assume all dimensions to be integer-valued. Uniformity of such encoding 
provides additional advantages for our techniques. 

The Cartesian product of the dimensional attribute domains forms the composite key space. The vector dependency $F$ then
maps a composite key to (a vector of) measures. Multidimensional database techniques are based on endowing the composite
key space with a space filling curve, so that each element of the space corresponds to a single point on the curve, 
and vice versa. There are multiple ways to choose such a curve; for our purposes we choose a class of curves called 
``generalized z-curve" (gz-curve), following \cite{orenstein:zcurves}. Each point on such curve is encoded with an 
integer that is derived from the values of the components of the composite key. Any query with point, range or set 
filters against the cube translates into a pattern search problem on the gz-curve. Such reduction is fairly standard 
and has been in use for many years in multidimensional databases, such as e.g. Oracle Essbase \cite{oracle:essbase}. 
Precise definitions and explanations will be given in the next section. 

We consider the following problems:

{\sc PROBLEM 1}:
Present a computer algorithm and an associated cost model, for efficiently retrieving elements matching a given pattern in the absence 
of additional indices. 

{\sc PROBLEM 2}: 
If data is partitioned by keys, present an appropriate parallelizable algorithm.

Grasshopper algorithms are applied after the described reduction of the OLAP problem to a pattern search problem in 
(composite) key-value space. They are very simple, and any database system based on a key-value store 
that supports a simple functional interface can take advantage of these algorithms. We have tested the algorithms
with a few such stores, both standalone and distributed, memory and disk based, including in particular Apache HBase \cite{opensource:hbase}. 
The results we have seen are consistent with the 
theory behind the Grasshopper algorithms. Before each query, the grasshopper takes a decision when to hop, based 
on the characteristics of the underlying key-value store and the query.

Particular cases of the algorithms - for certain encodings - translate into well known techniques. For example, 
if fact data is ordered by dimensional attributes, query processing is straightforward when leading attributes are filtered. 
Other cases, related to z-curves, have been considered in \cite{bayer:ubtree, markl:thesis, orenstein:zcurves}. However, 
to the best of our knowledge, Grasshopper algorithms have not been presented in the literature 
or implemented in relevant products.

The paper is organized as follows. Section \ref{sec:background} contains the necessary formalizations and reduction procedures. 
Related work is discussed. In section \ref{sec:algorithms} we describe Grasshopper strategy and provide a cost 
model. Then we provide Grasshopper algorithms for queries with point, range and set filters. In section \ref{sec:results} 
we present testing results and discuss them.

\section{Background and Related Work}\label{sec:background}

We provide the necessary background describing transformation of the OLAP fact data to key-value form (section \ref{sec:reduction}), 
discuss related work in section \ref{sec:related} and reformulate the problem as a pattern search problem in section \ref{sec:psp}.

\subsection{Reduction to key-value form} \label{sec:reduction}
Reduction from general cube schema to integer encoded dimensions is fairly common in the OLAP world. For the facts,
usage of composite keys is more typical in multidimensional OLAP (MOLAP). We outline the corresponding setup, in order to set the context
for further exposition. A similar setup is described in detail in Volker Markl's thesis \cite{markl:thesis}. Our algorithms do not depend
on particular ways of encoding dimensions.

In OLAP field, the main subject of study is a vector functional dependency $F: (D_1, ..., D_N) \rightarrow (M_1, ..., M_K)$. Here independent 
variables $D_i$ are called dimensions (dimensional attributes) and dependent variables $M_i$ are called measures. This dependency is augmented with additional, 
typically scalar, functional dependencies involving dimensional attributes, e.g. $City \rightarrow State$. Dependent attributes 
are called {\it higher level} dimensional attributes. They induce groupings within domains of the attributes they depend on
and, by virtue of that - aggregation operations on measures. Altogether dependencies are assumed to form a DAG.

In the assumed setup, all the dimensional attributes are encoded with integers. If the attribute is integer valued, it can be used without additional 
encoding. Otherwise an encoding dictionary is created. For attributes which are naturally ordered, encoding is required to preserve the order. 
Dense encoding (by consecutive integers) is preferred in most cases. How exactly encoding is organized is beyond the subject of this paper. 
For simplicity, we assume the cardinality of each dimensional attribute to be a power of 2. 

All dimensional dependencies are then expressed in encoded terms, often simply as arrays of integers, sometimes as graphs. 
It is supposed that both dictionaries and dependencies provide constant lookup time.

All the dimensional attributes that are of interest for analysis can participate in the formation of a {\it composite key}. Including higher level 
attributes in the key adds sparsity to the model, but often eliminates the need for joins at query time. 

Encoding with composite key transforms the data for functional dependency $F$ into key-value format. A {\it storage component} is responsible for
maintaining data in key-value format and retrieving relevant key-value pairs at query time. 

A simple query against the cube restricts certain attributes to a subset of their values and asks for data satisfying the restrictions. We study a class of filters 
on dimensional attributes, more precisely, point, range and set restrictions on the attributes' domains. 

When such query arrives, the system looks up the attribute values involved in the restriction against the dictionary and translates them into integers. These
integers are then used to form restrictions on the composite key; those are passed to the storage component for retrieving relevant key-value pairs, which are
aggregated, sorted, etc. as required by the query. Finally, dictionaries are used again to translate final results back to original attribute domains.

The above setup is common in many OLAP systems. Our algorithms fit in this picture as a vehicle for storage component to retrieve relevant key-value pairs quickly.

Let us explore composite key composition more closely. No matter how the composite key is produced, it provides integer encoding of all potential points in 
the key search space which is a Cartesian product of encoded attribute domains. Thus it provides (integer) parameterization for a space filling curve in
the search space. We restrict our attention here to a specific class of space filling curves, generalized z-curves (gz-curves for brevity) in the sense of \cite{orenstein:zcurves}. 
See also \cite{markl:thesis} for extensive exposition on the topic. Basically, the integer composite key in binary representation is built from bits of 
participating components' keys in a way that the order of bits of each component is preserved. This procedure produces keys of fixed length which is convenient for our purposes. 

\begin{figure}
%\centering
\hspace{-0.15in}
\includegraphics[width=5in]{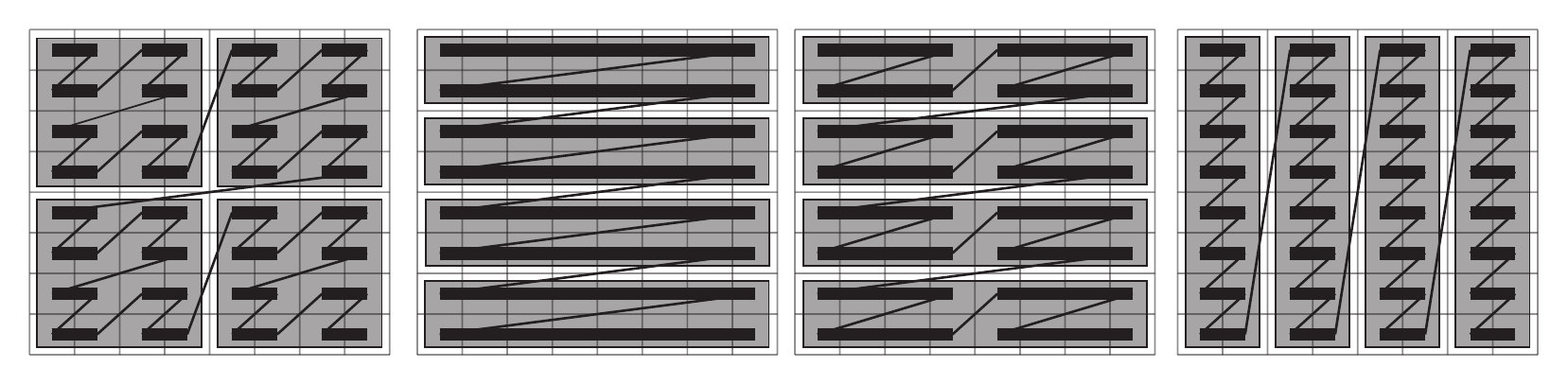}
\caption{Examples of gz-curves with bit orderings {\tt yxyxyx, yyyxxx, yyxyxx} and {\tt xxyyyx} respectively. Shaded areas are examples of fundamental regions (of order 4), 
to be described in section \ref{sec:geo}.}
\label{fig:curves}
\end{figure}

The shape of the gz-curve depends on the way component's bits are shuffled into the composite key. Figure \ref{fig:curves} shows a few possible shapes for two variables, where bits 
for horizontal and vertical dimensions are marked with {\tt x} and {\tt y} respectively. The first example is the classical ``isotropic" z-curve, and the second one is known as ``odometer curve" 
and corresponds to sorting keys by {\tt y}, then by {\tt x}. The latter case strongly favors one dimension over the other. It is very well known that answers 
to queries with filters on the leading dimension(s) of the odometer are located within a single contiguous segment of the curve, whereas for filters on the trailing dimension(s) 
they are scattered across the curve. Our algorithms become trivial in this case: the grasshopper either crawls all the time or hops every time, from one horizontal segment to another. 

\subsection{Related work}\label{sec:related}

Indexing techniques based on gz-curve have been presented in \cite{bayer:ubtree, markl:thesis, orenstein:zcurves} along with relevant data structures such as e.g. the UB-Tree. 
Besides the odometer ordering, the case of the isotropic curve has been well studied, especially for range filters on {\it all} of the dimensions 
(see e.g. \cite{bayer:integrating, tropf:rangesearch}). In the latter case, the bounding box of the query is often fairly densely covered by a single interval of the curve, 
and even full scan of that interval is very efficient. This is often exploited in geo-spatial applications. Our grasshopper typically does not have to jump in that case either. 

There are certainly plenty of similarities between our algorithms and UB-Tree methods of \cite{markl:thesis} and \cite{bayer:integrating}.
Efficiency of query processing is the key requirement for ad-hoc queries. Ability to perform all calculations directly in terms of the composite keys, without decomposing back
into dimensional coordinates, is crucial. In \cite{markl:thesis} such methods are briefly outlined for the case when  {\it all} dimensions have range or point restrictions. They use the specialized 
UB-Tree data structure that splits keys into regions (pages). The query is executed after determining which regions cover its bounding box. The number of such regions can often be quite high
(for incomplete range queries nearly all pages may qualify, for example) and determining all of them upfront can be quite lengthy, resulting in inefficiency. 

Knowing which pages to retrieve is extremely important for systems with disk storage because of high I/O costs. How the search is done within the page already in memory is also important,
because it allows controlling CPU costs. In the recent years, hardware has changed dramatically, and these days data can often completely fit into memory on one or more machines. 
As a result, for a particular storage organization, costs of certain operations can vary significantly. 

Taking that into account, we have decided to take a different stance. Since we cannot change the storage in many cases, we focus on scanning algorithms rather than data structures in 
the hope that they will be useful for many relevant systems. As a result, our algorithms, which do not determine the cover in advance and act in a streaming fashion, seem simpler to implement. 
Besides, our methods apply to any combination of point, range and set filters on any subset of dimensions, not necessarily on the full set. We can therefore argue that our
algorithms can improve performance of ad-hoc OLAP queries for any underlying key-value storage system that keeps data in the order of composite keys and supports certain simple operations. 
This is especially relevant in today's distributed storage systems, like e.g. HBase, where full scan is often the only option to answer queries. Of course, significant performance gains 
can be expected when the storage system efficiently supports the required operations. 
We discuss this in more detail when test results are presented. Basically, the grasshopper has in possession only certain characteristics of the storage, such as the ratio of 
sequential access and random access costs. Given a query, it computes a certain threshold beyond which it will jump if it encounters an appropriate ``obstacle" (unqualified key) 
while crawling. The threshold can be determined algebraically and explained geometrically. 

As already mentioned, we consider two problems. The first problem is very generic and does not require much from the storage system, except the ability to jump. The grasshopper does not 
know much about specifics of the storage. It may jump over keys within the same unit of storage, or it may jump landing on a different unit of storage. For the second problem, the grasshopper
can take advantage of the additional information provided by the storage, namely, the boundaries of the partitions by key intervals on the gz-curve. 
These partitions may correspond e.g. to pages of the UB-Tree or HBase regions, etc. Partitioning can be hierarchical, and is specific to storage. The grasshopper can then decide whether it 
needs to examine contents of the region or can skip it. In the case of the UB-Tree, this is essentially what the algorithms in \cite{markl:thesis, bayer:integrating} are doing. Our 
techniques apply actually to a more general class of {\it factorizable} partitions, to be defined later. Each partition can be processed in parallel, as is the case with HBase regions. 
Moreover, within a partition, we can typically reduce the dimensionality of the problem and operate directly on the reduced (factorized) keys, without the need to restore the original 
keys which is a costly operation. More on this in the upcoming sections.

\subsection{Pattern Search Problems}\label{sec:psp}

In this section we describe our problems as pattern search problems on the gz-curve.

Any point restriction on one or more of the attributes means fixing a pattern of bits in the key, so the query problem translates into a fixed pattern search problem (PSP) on a set of keys. 
Similarly, range and set restrictions result in patterns, albeit more complex. Definitions provided below are aimed at expressing everything in the pattern search related terminology.

Let $n$ be the total number of bits in the composite key. Consider the space of all keys $S$ as ${\mathbb Z}_2^n$, an $n$-dimensional linear space over the group of residues 
${\mathbb Z}_2 =  \{0,1\}$. 
The bits form an ordered basis $e_1, \ldots , e_n$ in $S$, and elements of $S$ are ordered lexicographically by coefficients (which trivially coincides with the order of integers).

Define {\it mask} to be an operator of projection onto a $d$-dimensional coordinate linear subspace of $S$.  Given $d$ basis 
vectors $e_{i_1},\ldots ,e_{i_d}$, such operator $m$ simply masks out the remaining $n-d$ coordinates.
Denote by $S(m)$ the subspace onto which the mask $m$ projects. 

Two or more masks are called {\it disjoint}, if the subspaces onto which they project are pairwise disjoint, i.e. do not have any 
common basis elements.

To emphasize similarity with bit masking, we will use the notation $x \& m$ for $m(x)$. We will also use notation $p | q$ instead of the sum of vectors $p$ and $q$ belonging to two disjoint 
subspaces, and similarly, for masks. 

For the case of gz-curves, to each dimensional attribute $D$ corresponds a mask $m_D$ that defines its bit positions in the composite key. Applying the mask to the composite key retrieves 
the contributing value of $D$. Obviously, masks corresponding to different dimensional attributes are disjoint.

Let $A$ be a subset of $S$ representing composite keys of the cube fact data. Any query against the cube with point filter $D = p$ on attribute $D$ translates into a pattern search 
problem (PSP): find all $x \in A$ such that $x \& m_D = p$. Queries with point filters $D_i = p_i$ on multiple attributes $D_i$, translate into a similar problem of 
finding solutions to $x \& m = p$, for the union $m$ of attribute masks and union $p$ of corresponding patterns.

It is convenient to consider {\it any} mask on $S$ as corresponding to some ``virtual" attribute. Thus a query with multiple point filters is equivalent to a query with a single point 
filter on an appropriate virtual attribute.

A query with range filter $D \in [a, b]$ in this setting also translates into a PSP: find all $x \in A$ such that $x \& m_D \in [a, b]$. However, unlike the point case, 
combining two or more such queries into a single similarly expressed query against some virtual attribute is generally impossible. So we will have to look for elements that 
simultaneously satisfy some number of patterns. 

Set queries are transformed in a similar fashion. Given a set $E=\{a_1, \ldots , a_N\}$, the filter $D \in E$ corresponds to the PSP $x \& m_D \in E$. Set filters for
multiple attributes can be combined into a similarly expressed query against some virtual attribute, but since the resulting restriction set is a Cartesian product of coordinate restrictions,
its cardinality may be too large for practical purposes, so multi-pattern search is used for them as well. Obviously, any solution to the search condition also satisfies a range restriction
$x \& m_D \in [\min(E), \max(E)]$.

To summarize, we consider pattern search problems (PSP) with restrictions of the following kinds: $x \& m = p$ (P), $x \& m \in [a,b]$ (R), $x \& m \in \{a_1, \ldots, a_N\}$ (S). 

A brute force solution to the pattern search problem on a set $A \subset S$ is achieved via checking pattern restrictions on each element of $A$ (full scan). A solution to the pattern 
search problem will be called {\it efficient}, if on average it is faster than the brute force solution and is never slower than that. 
To comply with ad-hoc query requirements, the average here is taken with respect to a set of random pattern restrictions on any fixed combination of the appropriate number of 
attribute restrictions and then over all such combinations. According to this definition, an efficient algorithm is allowed to lose to the full scan on some pattern, 
but is not allowed to lose on average. We provide cost estimates explaining why our algorithms are efficient and confirm this by experiments.

A set $X\subset S$ is called {\it factorizable} if it can be represented as a Cartesian product of at least two subsets of $S$. Besides $S$ itself, the set of all its elements
satisfying a restriction of kind (P) is clearly factorizable. This is also true for restrictions of kind (R) and (S), when $S(m) \neq S$.%, since $S(\ttilde m)$ is then a non-trivial factor. 
Examples of factorizable subsets include intervals with common prefix or sets with common pattern.

Let $\{S_j\}$ be a partition of $S$ into factorizable subsets (perhaps each with its own factors). The induced partition of any $A\subset S$ by sets $A_j=A\cap S_j$ is 
also called {\it factorizable}. 

Our algorithms get additional advantage when dealing with factorizable partitions, in particular with partitions by key intervals or by sets with common pattern. They become especially
efficient when the underlying storage implements prefix or common pattern compression.

The problems Grasshopper algorithms are intended for can now be more precisely formulated. 

{\sc PROBLEM 1}: Let $m_1, \ldots, m_k$ be an arbitrary collection of disjoint masks on space $S$. Let, for each of these masks, a pattern restriction of kind (P), (R) or (S) be given. 
Further, let $A$ be an arbitrary subset of $S$. Find an efficient solution to pattern search problem on $A$. 

{\sc PROBLEM 2}: In the same setting, when $A$ is partitioned in a factorizable manner, present an efficient parallelizable algorithm.

\section{Analysis and Algorithms} \label{sec:algorithms}

In this section we first present the Grasshopper strategy for solving our Problem 1. 
In section \ref{sec:masks} we develop some notation and definitions needed for its implementation. 
Then, for each kind of pattern restrictions, we introduce the algorithm that helps to implement this strategy. Grasshopper algorithms use geometric properties of PSP solution 
loci on the gz-curve, so we outline them before describing the algorithms in each case. Geometric properties for point PSP are presented in section \ref{sec:geo}. Point matchers
are introduced in section \ref{sec:point}. We then explain in section \ref{sec:partitioned}, how to handle our Problem 2 for point restrictions. 
In sections \ref{sec:range} and \ref{sec:set} we introduce range and set matchers respectively. In section \ref{sec:combine} we outline how queries with 
multiple restrictions of various kinds are handled.

\subsection{Grasshopper strategy}

Here we present Grasshopper strategy for scanning data in search of particular patterns. It is designed to avoid doing a full scan of the data by intelligently combining sequential 
crawling with jumps over large portions of irrelevant keys

Desire to have algorithms applicable to any underlying data structure led us to split powers between the devices used in the pattern search: the data store and the pattern matcher. We
will formulate our algorithms using these concepts. The data store is typically not in our control; pattern matchers together with the Grasshopper strategy make up the Grasshopper 
algorithms.

A key-value store is assumed to contain key-value pairs whose keys are elements of $A\subset S$ and to be aware of the key ordering. In terms of its capabilities, a data store will be 
called {\it basic} if it supports the following operations:
\begin{description} \itemsep1pt \parskip0pt
\item [Get:] given a key $x\in A$, retrieve the appropriate value.
\item [Scan:] given a key $x \in A$, retrieve the next key in $A$. 
\item [Seek:] given a key $x \in S$, retrieve the next key in $A$ larger than or equal to $x$.
\end{description}
We also suppose that statistics of $A$ (such as cardinality, first and last key) are available at negligible cost.

A {\it partitioned} data store is supposed to be able to provide partitioning criteria and to possess the above basic capabilities for each element of the partition.

Besides the data store, the other player in the game is a {\it matcher}. The matcher is designed to assist with pattern search assuming the following functionality:
\begin{description} \itemsep1pt \parskip0pt
\item [Match:] given $x\in S$, tell whether $x$ satisfies the given pattern restrictions. 
\item [Mismatch:] given $x\in S$, return 0 if $x$ satisfies the given pattern restrictions or the (signed) position of the highest bit in $x$ responsible for mismatch, with sign indicating 
whether mismatch is from above or from below.
\item [Hint:] given an element $x \in S$ with its mismatch $y$, suggest the next element $h\in S, h > x$, that can theoretically satisfy the pattern restrictions.
\end{description}

The roles are quite distinct: the store knows everything about the set $A$, but nothing about the masks and patterns; for the matcher it is the other way round.
All the algorithms follow the same pattern, but matcher variations for different cases of pattern problems are quite different.

A race over the data store is announced. Participants must collect data matching a given set of patterns, into a bag. At the start of the race, there is a crawler, 
a frog and a grasshopper. Each one is given a matcher. Using the matcher, they can compute the theoretical query bounding interval $[PSP_{min}, PSP_{max}]$ 
on $S$ and intersect it with the interval $[\min(A), \max(A)]$ to obtain the actual bounding interval $[a,b]$.

The crawler's strategy is sequential scan:

\begin {tt}
bag = $\emptyset$; $x = a;$ 

while $x \le b\ \{$

\ \ if ${\tt Match} (x)$, add $(x, {\tt Get}(x))$ to bag; 

\ \ $x = {\tt Scan} (x);$ 

\}
\end{tt}
		
The frog's strategy is jumping as soon as possible: 

\begin{tt}
bag = $\emptyset$; $x = a;$ 

while $x \le b\ \{$

\ \ y = ${\tt Mismatch} (x);$

\ \ if $y = 0$, add $(x, {\tt Get}(x))$ to bag, $x = {\tt Scan} (x)$; 

\ \ else $x = {\tt Seek} ({\tt Hint} (x, y))$;

\}
\end{tt}

The grasshopper's strategy is to jump only when the absolute value of the mismatch is above a certain threshold $t$; otherwise crawl:		

\begin{tt}
bag = $\emptyset$; $x = a;$ 

while $x \le b\ \{$

\ \ y = ${\tt Mismatch} (x);$

\ \ if $y = 0$, add $(x, {\tt Get}(x))$ to bag, $x = {\tt Scan} (x)$;

\ \ else if $|y| <= t, \ x = {\tt Scan} (x)$;

\ \ else\ $x = {\tt Seek} ({\tt Hint} (x, y))$;

\}
\end{tt}

When the {\tt Hint} operation has nothing to suggest, it returns $\infty$ and the corresponding loop terminates. 

Some modifications on the grasshopper strategy will be outlined later on.

The following simple cost model can be suggested. First, observe that all three racers are doing the same number of ${\tt Scan}$ and ${\tt Get}$ operations for those
$x$ that do match the PSP restrictions. We can exclude these from the cost estimates. The only difference for matching elements 
may be in the cost of ${\tt Match}$ and ${\tt Mismatch}$ operations. 

For those elements, that do not solve the PSP, the crawler performs ${\tt Match}$ and ${\tt Scan}$ operations, the frog performs ${\tt Mismatch}$, ${\tt Hint}$ and ${\tt Seek}$
operations, and the grasshopper sometimes does the same as the crawler and sometimes - the same as the frog. 

Let us assume that matcher's operations take negligible time compared to the data store's operations (this is often true in reality). So the essential costs to compare are: 
\begin{description} \itemsep1pt \parskip0pt
\item [Crawler:] $N_0 \cdot cost ({\tt Scan})$;
\item [Frog:] $N_1 \cdot cost ({\tt Seek})$;
\item [Grasshopper:] $N_2 \cdot cost ({\tt Seek}) + N_3 \cdot cost ({\tt Scan})$.
\end{description}

Here $N_0$ is the number of mismatched elements, $N_1$ is the number of frog's jumps, $N_2$ is the number of grasshopper's jumps, and $N_3$ is the number of times grasshopper continues to crawl. These values are functions of $m$ and $p$.

Recall that, by our definition of efficiency, we need to compare average costs. Fix the PSP mask $m$, and let $\overline {N_i}$ be the sum of the corresponding $N_i$ over all $2^d$ possible patterns. Let $A[m,p]$ be the corresponding PSP solution.

Let $R = cost ({\tt Scan}) / cost ({\tt Seek})$. $R$ is a property of the data store that can be experimentally determined (it may depend on the data set). When data is physically stored sorted, usually $R<1$, but if navigation is via an index, it may happen that $R=1$. For in-memory data store, $R$ is typically closer to 1 than for disk-based one.

It is clear that the frog will finish (on average) ahead of the crawler if $\overline{N_1} < \overline{N_0} \cdot R$. 

The term $\overline {N_0}$ is the sum of cardinalities of $A\setminus A[m,p]$, and since $A[m,p]$ form a disjoint cover of $A$, we obtain $\overline{N_0} = (2^d-1) \cdot card(A)$. 

So the frog really hopes that

\begin{equation}\itemsep1pt \parskip0pt
\overline {N_1} < R\cdot card(A)\cdot (2^d - 1) \label {N1}
\end{equation}

Obviously, the opposite inequality is in favor of the crawler. 

The right hand side of (\ref{N1}) does not depend on the geometry of the mask(s), i.e. on the way the attributes participate in the key composition. 
By contrast, $N_1$ is heavily dependent on the mask. 

If the grasshopper determines ahead of the race that the frog is guaranteed to win over the crawler, it can solidarize with the frog by deciding to use the threshold value $t=0$. 
However there are cases when the frog definitely loses to the crawler. For example, if the mask consists only of the first (most junior) bit, then theoretically every second point of $S$ 
solves the PSP. The matcher cannot propose anything better than jumping exactly to the next point, definitely a losing strategy.

Well, in the worst case the grasshopper can solidarize with the crawler by deciding to use the threshold value $t=n$ prohibiting any jumps. But the grasshopper cannot always 
solidarize with the crawler, because then its strategy will not meet the ``efficiency" criteria. And the grasshopper must make its decision before the race! So it must be able to at 
least verify if (\ref{N1}) holds.

In order to help the grasshopper to make the right decision, we need to examine the problem more closely.

\subsection{Masks and Patterns} \label{sec:masks}

In this section we develop additional notation and terminology in order to formulate our solutions.

For any mask $m$, there exists a complementary mask (co-mask) $\ttilde m$, that projects onto the remaining $n-d$ coordinates. Certainly, any $n$-dimensional vector $x$ can be 
restored from its projections onto $S(m)$ and $S(\ttilde m)$, i.e. $x = m(x) | \ttilde m(x)$.

It is clear that co-mask definition can be extended to complement a set of masks $m_1,\ldots,m_k$ (just pick a subspace orthogonal to the span of $S(m_i)$).

We say that mask $m$ is {\it covered} by masks $m_1,\ldots,m_k$ if both $m$ and $m_1, \ldots,m_k$ jointly have the same complement $m'$. A cover is a {\it partition}
if spaces $S(m_i)$ do not have common basis vectors.

Let mask $m$ be projecting onto bits $e_{i_1},\ldots,e_{i_d}$ in $S(m)$, listed in ascending order. 
Define $tail (m) = i_1-1$ and $head (m) = i_d$. Mask $m$ is called {\it contiguous} if it projects onto adjacent bits, or, equivalently, $head (m) = tail (m) + d$.

For a mask $m$ and some element $e_i$ of the basis of $S$, denote by $m_{>i}$, $m_{=i}$ and $m_{<i}$ projections of $m$ onto basis vectors $e_{i+1},\ldots, e_n$, 
onto $e_i$ and onto basis vectors  $e_1,\ldots,e_{i-1}$ respectively. Thus $m = m_{>i} | m_{=i}  | m_{<i}$, with similar relationships for the projection spaces. One or more 
of the projections may be empty. Similarly, any pattern $p$ can be decomposed as $p = p_{>i} | p_{=i}  | p_{<i}$.

An element of a subspace $T\subset S$ all of whose coordinates are 0 (1) is denoted $0_T$ ($1_T$), and we will write $0_m$ instead of $0_{S(m)}$. 

Define a partial order on the set of masks as $m_1 > m_2 \Leftrightarrow tail (m_1) >= head (m_2)$. 
Among partitions of mask $m$ by contiguous masks, {\it canonical partition} of $m$ is the one with the smallest number of parts. 
We will always enumerate them in descending order, from senior bits to junior ones.

The smallest element $PSP_{min}$ matching a fixed pattern $p$ in $S$ is of the form $0_{\ttilde m} | p$ and the largest such element - $PSP_{max}$ is $1_{\ttilde m} | p$.
These elements form the bounding interval for the fixed pattern search problem. Although they depend on $p$, their difference, $spread (m, PSP)$, does not: 
$spread (m, PSP) = 1_{\ttilde m} | 0_m.$

For any range or set pattern restriction, with minimal element $a$ and maximal element $b$, we have $PSP_{min} = 0_{\ttilde m} | a$ and $PSP_{max} = 1_{\ttilde m} | b$. 
For contiguous masks, the spread depends only on the difference $b-a$, however this is not true in the general case.

\subsection{Geometrical properties of the point PSP locus on gz-curve}\label{sec:geo}

In this section we obtain certain geometric properties of the point PSP locus on the gz-curve. 

Recall that the gz-curve is used as a space filling curve for the Cartesian product $T$ of $N$ attribute domains of integers. The cardinality of each domain $D_i$ is a power of 2, 
and the way it 
participates in forming the element of the gz-curve is expressed by the domain mask $m_{D_i}$. In the previous section, the space filling curve was algebraically 
expressed as an $n$-dimensional space, $S$, with $n$ being the total number of bits of all domains. How do $T$ and $S$ relate to each other geometrically?

As the gz-curve traverses the space $T$, one can identify {\it fundamental regions} $T^r$ of order $r$ for $r=0,\ldots, n$ as the rectangular boxes with volume $2^{r}$ corresponding 
to intervals of the gz-curve (each $T^0$ region is a single point, and $T^n = T$). The ends of the intervals are aligned with the corresponding power of 2. 
See figure \ref{fig:curves} for an example.
Each fundamental region contains fundamental regions of lower orders. All the regions of given order $r$ are replicas of each other, the shape of the gz-curve in them is the same 
and their number in $T$ is $2^{n+1-r}$. When no confusion arises, corresponding intervals on the gz-curve will also be called fundamental. 

Solution locus of any PSP on a gz-curve consists of certain intervals (clusters), which in some cases degenerate to a point. We will use the term {\it lacuna} for the gap {\it between} the clusters
(excluding gaps at the ends of the curve). 

We would need to compute certain quantities characterizing the locus of any point PSP: cluster count, cluster lengths, total lacunae length, and individual lacunae lengths. 

\newtheorem{proposition}{Proposition}
\begin{proposition}\label{prop:points}
Let $m$ be an arbitrary mask projecting onto $d$ dimensions, and let $\{m_i\}$ be its canonical partition. Let $x\& m = p$ be a point PSP. 

Then the locus of the PSP consists of $2^{n-d - tail(m)}$ intervals of length $2^{tail (m)}$ each, separated by lacunae of total length
$spread (m, PSP) - 2^{n-d}$. The spread can be calculated as $2^n - \overline {m}$ where $\overline {m} = 1_m | 0_{\ttilde m}$. 
Individual lacunae lengths are partial sums 
\begin{equation}\label{points:lacunae} 
\Sigma_j = \sum_{i\ge j} [2^{head(m_i)} - 2^{tail (m_i)}].
\end {equation}
\end{proposition}

\begin{figure}
\centering
\includegraphics[width=3.2in,height=0.8in]{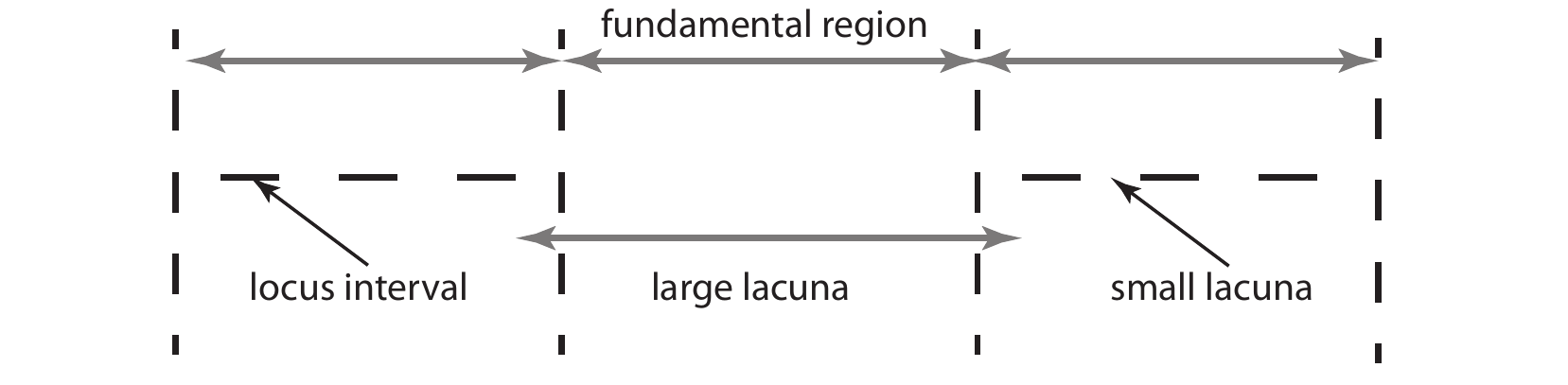}
\caption{Structure of the locus for point PSP.}
\label{fig:lacunae}
\end{figure}

We briefly outline the proof which follows by induction on the number of the elements in the canonical partition of the mask. 
For a contiguous mask with $d$ bits, it is easy to see that only one out of $2^d$ adjacent intervals of size $2^{tail(m)}$ 
within any fundamental region $T^{head(m)}$ 
qualifies for the given fixed pattern restriction with mask $m$ and is also a fundamental region $T^{tail(m)}$. There are $2^{n-head(m)}$ such regions. 
Thus the locus of points on the gz-curve satisfying the restriction consists of $2^{n-head(m)}$ clusters of length $2^{tail(m)}$ 
separated by lacunae of length $2^{head(m)} - 2^{tail(m)}$. See figure \ref{fig:lacunae} for an illustration.

The step of the induction follows by repeating the base argument within each of the fundamental regions $T^{tail(m_k)}$ 
(previously identified matching cluster) considered instead of $S$ for the next component of the mask. 
This time however we need to take into account the gaps at the edges of the regions. These additional gaps have the length  
$2^{head(m_{k+1})} - 2^{tail(m_{k+1})}$ which yields (\ref{points:lacunae}).

Note. It is easy to see that partial sums $\Sigma_j$ in (\ref{points:lacunae}) have the following bounds:
\begin{equation}\label{points:lacunae-bounds}
0 < 2^{head(m_j)} - \Sigma_j \le 2^{tail(m_j)} \le 2^{head(m_j)-1}.
\end{equation}

\subsection{Grasshopper algorithm for point queries}\label{sec:point}

In this section we describe Grasshopper algorithm for the point matcher.

Let $m$ be some mask projecting onto $d$ dimensions, and let $p$ be an element of $S(m)$. For a subset $A\subset S$, consider the fixed pattern search problem 
$PSP (m, p)$, that is, finding all elements $x\in A$ such that $x\& m=p$. Any mask partition $\{m_i\}$ also induces pattern partition $\{p_i\}$. 
An element of $A$ matches $p$ on $m$ if and only if it matches $p_i$ on $m_i$ for each $i$. 

Here is how matcher operations for this PSP are defined. 
For the {\tt Mismatch} operation, the matcher works by examining $PSP(m_i, p_i)$, $i=1,\ldots$, one at a time. If $x\& m_i \neq p_i$, let $e_j$ be the most senior bit, 
on which the sides disagree. The matcher returns $j$ if $x\& m_i > p_i$ and $-j$ if $x\& m_i < p_i$. If $x\& m_i = p_i$, the matcher proceeds on to $PSP(m_{i+1}, p_{i+1})$, 
and so on. If no mismatch is detected, the matcher returns 0. 

Let $I$ be the identity mask on $S$, i.e. the mask projecting onto entire $S$.

For the {\tt Hint} operation, given an element $x\in S$ and mismatch position $j$, the matcher acts as follows. 

If mismatch is negative and indicates mismatch at bit $j$, the matcher returns $hint(x, j) = x_{I_{>j}} | 1_{I_{=j}} | p_{m_{<j}} | 0_{\ttilde m_{<j}}$. 
The highest bit position that changes is $j$. Geometrically this means that the point $x$ belongs to some fundamental region $T^{j-1}$, and by changing the bit, 
we are placing the result into the next such region. Since none of the bits above $j$ is changed, 
we are staying within the same fundamental region $T^j$ that contained $x$. 

If the mismatch is positive, geometric meaning of the operation is similar, but the next fundamental region $T^{j-1}$ intersecting with the PSP locus, is located in a different 
fundamental region of higher order than the one containing $x$. In order to find such region, we need to determine the ``growth point" $g$ which is the smallest position 
above $j$ of an unset (0) bit in $x\& (\ttilde m)_{>j}$. If such position does not exist, the search is over ($\infty$ returned), 
otherwise the value $hint(x, g)$ is returned.

We need to estimate from above the total number of times when the frog jumps, $\overline{N_1}$. Note that the jump occurs only when a mismatch is detected, i.e. when $x$ belongs to some lacuna. After the jump,
the frog lands on the next cluster. Clearly the number of jumps cannot exceed the number of lacunae, which, by proposition \ref{prop:points}, is $2^{n-d - tail(m)} - 1$ and does not 
depend on $p$. Hence $$\overline{N_1} \le 2^d \cdot (2^{n-d - tail(m)} - 1).$$ If the right hand side is less than  
the right hand side of (\ref{N1}), the frog finishes (on average) ahead of the crawler. Such estimate obviously holds for some masks, for example, for contiguous headless masks, 
since for those $n = d+tail(m),$ and there are no lacunae. Rewrite the condition (\ref{N1}) as 

\begin{equation}\label{R1}
R > R_1(m, A) = (2^{n-d-tail(m)}-1) / [card(A)\cdot (1 - 2^{-d})].
\end{equation} 

The above estimate is instrumental in the {\it dense} case, when the cardinality of $A$ is comparable to the number of lacunae. In the {\it sparse} case, when we have much less points than lacunae, a different estimate is needed. Consider the probability distribution $P$ of points of $A$ with respect to fundamental regions $T^{tail(m)}$ and let $P_i=P(T^{tail(m)}_i)$. The average number of points per region is ${\overline P} = 2^{-(n-tail(m))}.$ For a fixed pattern, each region may participate in at most one lacuna. 

For a given mask $m$, define {\it region co-frequency} $k_i$ as the number of times the $i$-th region participates in lacunae across all patterns, $0\le k_i\le 2^d-1$. 
The co-frequencies are easy to determine given the mask. For example, for contiguous masks, co-frequencies are $0, 1 ... 2^d-1...2^d-1, 2^d-2 ... 0$. For non-contiguous masks, co-frequencies consist of several series of similar kind with various exponents up to d.

The total number of points $M$ in all lacunae across all patterns is $card(A) \cdot \sum k_i P_i$. Since $\overline{N_1} \le M,$ condition (\ref{N1}) can be rewritten as $$card(A) \cdot \sum k_i P_i < R \cdot card(A) \cdot (2^d-1),$$ or,

\begin{equation}\label{R2}
R > R_2(m, A) = \sum k_i P_i /(2^d -1). 
\end{equation}

Observe that $R_2(m, A) \le 1.$

We have arrived at the following conclusion:

\begin{proposition}\label{prop:frog}  
Let $m$ be a mask projecting onto $d$ dimensions (bits) of $S$, and let $A$ be a nonempty subset of $S$. Let $T^{tail(m)}$ be fundamental regions with co-frequencies $\{k_i\}$, and let  $P$ $(\{P_i\})$ be the distribution of $A$ with respect to those fundamental regions. Let $R_1 = R_1(m, A)$ and $R_2=R_2(m, A)$ be defined by (\ref{R1}) and (\ref{R2}) respectively. 

If the scan to seek ratio of the data store satisfies the estimate $$R > \min (R_1, R_2),$$ then the frog strategy is more efficient than the crawler strategy. 
\end{proposition}

The condition $R>R_2$ would be meaningful if $R_2 < 1.$ The value of $\sum k_i P_i$ might not be straightforward to compute, let us examine particular situations when the estimates can be simplified. 

First, examine the case when the distribution is uniform, i.e. $P_i = {\overline P}.$ Note that $\sum k_i$ is proportional (with coefficient $2^{-tail(m)}$) to the length of all lacunae for all patterns. By proposition \ref{prop:points}, the length of lacunae for each pattern is $$spread (m, PSP) - 2^{n-d}= 2^n - \overline{m} -2^{n-d}.$$ Hence, over all patterns,  
$$\sum k_i = 2^{d-tail(m)} (2^n - \overline{m} -2^{n-d}).$$ Observe that the minimum value 
of $\overline{m}$ is achieved when $m$ is a contiguous tailless mask projecting onto $d$ most junior bits. Thus $\min(\overline{m})=2^d - 1$, and 
$$2^n - \overline{m} -2^{n-d} = 2^n(1 - 2^{-n}\overline{m} - 2^{-d}) \le 2^n(1 - 2^{-n}(2^d-1) - 2^{-d}) = 2^n(1-2^{d-n})(1-2^{-d}).$$ 
It follows that $$\sum k_i \le 2^{n-tail(m)} (1-2^{d-n})(2^d-1)$$ 
and
$$R_2 \le \overline{P}\cdot 2^{(n-tail(m))} \cdot (1-2^{d-n}) = 1 - 2^{d-n} < 1.$$ 

Clearly, the same would be true when distribution is sufficiently close to uniform. Observe that, for a given $\epsilon,$
$$\sum_i k_i P_i = \sum_{P_i - {\overline P} <\epsilon{\overline P}} k_i P_i + \sum_{P_i - {\overline P}\ge\epsilon{\overline P}} k_i P_i.$$ The first term can be estimated as
$$\sum_{P_i - {\overline P}<\epsilon{\overline P}} k_i P_i < {\overline P}\cdot (1+\epsilon) \cdot \sum k_i \le {\overline P}\cdot 2^{n-t} (1+\epsilon) (2^d-1) (1 - 2^{d-n}).$$ 
If  $\sigma^2 = 2^{-(n-tail(m))} \sum (P_i - {\overline P})^2$ is the variance of $\{P_i\}$, it follows that the number of terms in the sum for which $P_i - {\overline P}\ge\epsilon{\overline P},$ cannot exceed $2^{n-tail(m)} \sigma^2 / (\epsilon {\overline P})^2 = 2^{3(n-tail(m))} \sigma^2 / \epsilon^2.$ Hence  
$$ \sum_{P_i - {\overline P}\ge \epsilon{\overline P}} k_i P_i \le \max k_i \cdot 2^{3(n-tail(m))} \sigma^2 / \epsilon^2 = {2^{3(n-tail(m))}\sigma^2(2^d-1)\over {\epsilon^2}}.$$ 
%and for the second term, by Chebyshev inequality, we have
%$$ \sum_{P_i - {\overline P}\ge \epsilon{\overline P}} k_i P_i \le \max k_i \cdot  P(\{P_i - {\overline P}\ge \epsilon{\overline P}\}) \le \max k_i \cdot {\sigma^2 \over {\epsilon^2}{\overline P}^2} %= {\sigma^2(2^d-1)\over {\epsilon^2}{\overline P}^2},$$ where $\sigma^2$ is the variance of $\{P_i\}.$ 
Then 
\begin{equation}\label{R2estimate}
%R_2 < (1+\epsilon)(1 - 2^{d-n}) + {\sigma^2 \over {\epsilon^2}{\overline P}^2} =  (1+\epsilon)(1 - 2^{d-n}) + {2^{2(n-t)}\sigma^2 \over {\epsilon^2}}.
R_2 < (1+\epsilon)(1 - 2^{d-n}) + {2^{3(n-t)}\sigma^2 \over {\epsilon^2}}.
\end{equation}

First choose an $\epsilon$ that would minimize the value of the right hand side. The right hand side has the form $a + a\epsilon + b^2 /\epsilon^2,$ with $a=1 - 2^{d-n}$ and %$b= 2^{n-tail(m)}\sigma$
$b= 2^{3/2(n-tail(m))}\sigma$.  Taking derivative to find the minimum, one obtains $a - 2b^2/\epsilon^3 =0$ or $\epsilon = (2b^2/a)^{1/3},$ which yields the minimum value of $$a + (2(ba)^2)^{1/3} + (2(ba)^2)^{1/3}/2= a + 1.5 \cdot (2(ba)^2)^{1/3}.$$ Since $a=1 - 2^{d-n}$, in order for the above expression to be less than 1, we need to have $$1.5 \cdot (2(ba)^2)^{1/3} < 2^{d-n},$$ or 
$$b < 2/a \cdot (2^{(d-n)} / 3)^{3/2}.$$ 

Substitute the expressions for $a$ and $b$ to see that the last inequality is equivalent to 
\begin{equation}\label{sigma0}
%\sigma < \sigma_0 (m) = {2^{1 - (n-tail(m)) + 1.5(d-n)} \over \sqrt{27} (1 - 2^{d-n})} = {2^{1 - 2.5n + tail(m) + 1.5d} \over \sqrt{27} (1 - 2^{d-n})}.
\sigma < \sigma_0 (m) = {2^{1 - 1.5(n-tail(m)) + 1.5(d-n)} \over \sqrt{27} (1 - 2^{d-n})} = {2^{1 - 3n + 1.5tail(m) + 1.5d} \over \sqrt{27} (1 - 2^{d-n})}.
\end{equation}
The right hand side of (\ref{R2estimate}) is
\begin{equation}\label{R2prime}
%R_2' = 1-2^{d-n} + 1.5\cdot (2^{2(n-tail(m))+1}(1-2^{d-n})^2\sigma^2)^{1/3}.
R_2' = R_2'(m) = 1-2^{d-n} + 1.5\cdot (2^{3(n-tail(m))+1}(1-2^{d-n})^2\sigma^2)^{1/3}.
\end{equation}

Sufficient conditions are summarized in the following 
\begin{proposition}\label{prop:sufficient}
Let $m$ be a mask projecting onto $d$ dimensions (bits) of $S$, and let $A$ be a nonempty subset of $S$. Let $\sigma,$ the standard deviation of the distribution of $A$ with respect to fundamental regions $T^{tail(m)}$ satisfy (\ref{sigma0}). Let quantities $R_1$ and $R_2'$ be defined by (\ref{R1}) and (\ref{R2prime}) respectively.

If the scan to seek ratio of the data store satisfies the estimate $$R > \min (R_1, R_2'),$$ then the frog strategy is more efficient than the crawler strategy. 
\end{proposition}

The grasshopper is clever enough to verify the conditions of Proposition \ref{prop:frog} (or \ref{prop:sufficient}) and, if they hold, set its strategic threshold to 0. The grasshopper will then arrive with the frog, ahead of the crawler.

Proposition \ref{prop:sufficient} can be interpreted in the following manner: if the mask has sufficiently high tail, the dense case condition kicks in, and the relatively small number of lacunae awards victory to the frog and grasshopper; if the mask has low tail, the sparse case condition must be checked, and the victory is guaranteed when the distribution over $T^{tail(m)}$ is close to uniform. For the frog, the chance of winning is limited to these cases. For the grasshopper, however, there are still opportunities to win over both. 

The idea is, even when the mask tail is low, but there are contiguous components with sufficiently high tails, setting the right strategy threshold would allow the grasshopper to win over the crawler because it jumps over large lacunae and to win over the frog because it scans small lacunae. Let us examine the situation in more detail. 

Grasshopper jumps when the matcher detects mismatch exceeding the threshold. Mismatch value for a non-qualified point carries 
information about the size of its lacuna, and the threshold $t\ge tail(m)$ is designed to ensure that only large lacunae trigger the jumps. 
Let $d_t$ be the number of bits in the mask above $t$. For a given pattern, PSP locus is contained in only $2^{n-t-d_t}$ of 
$2^{n-t}$ fundamental regions $T^t$ ({\it crawl regions}). The regions between crawl regions are {\it jump regions}. Geometric interpretation of
the grasshopper strategy is to crawl when the current point is within a crawl region and to jump if it belongs to a jump region. Adjacent jump regions form $2^{n-t-d_t}-1$ lacunae, and grasshopper jumps over the entire lacuna in one hop. 

First, consider competition against the crawler. Within smaller size lacunae both strategies perform scans, so crawl regions have no influence on the competition. Recall that lacuna sizes correspond to contiguous components of the mask. Removing components below the threshold would only affect the PSP locus in the crawl regions, hence trailing components
can be safely dropped from consideration without affecting the outcome of the competition between the crawler and the grasshopper. So assume that all components of the mask lie at or above $t$. Under this assumption, $d_t = d, N_2 = N_1, N_3 = 0$, and both frog and grasshopper strategies coincide.

The total number $J$ of all jump lacunae for all patterns is $2^d\cdot (2^{n-t-d}-1) < 2^{n-t}$, and $\overline {N_1}\le J$ . Rewrite the estimate (\ref{N1}) as 
$$2^{n-t} \le card(A)\cdot R \le card(A)\cdot R\cdot (2^d-1)$$ or, $$t\ge t(m, A) = n - \log_2 (card(A)\cdot R).$$ The value $t(m, A)$ can thus serve as threshold. 
This argument is similar to the one used to define $R_1(m, A)$.

To summarize, we have arrived at the following result:

\begin{proposition}\label{prop:grasshopper}  
Let $m$ be a mask projecting onto $d$ dimensions (bits) of $S$ with canonical partition $\{m_i\}$ and let $A$ be a nonempty subset of $S$. 
Let, further, $R$ be the scan-to-seek ratio of the data store, and let $$t(m, A) = n - \log_2 (card(A)\cdot R).$$
Then grasshopper strategy with threshold $t(m, A)$ is more efficient than the crawler strategy. 
\end{proposition}

Another way to arrive at a similar threshold value is to observe that lacunae must be sufficiently long to contain a large enough number $X$ of elements of $A$. 
The grasshopper will skip over $X-1$ of them as soon as it stumbles upon the first one. The crawler, however, 
would have to visit them all. Hence the difference between crawler and grasshopper strategy costs would amount to comparing $X\cdot \overline{N_2}$ and $\overline{N_2} / R$. Grasshopper wins when $X > 1/R$.

Let $j_0$ be the maximal value of all such $j$ for which the partial sum $\Sigma_j$ in (\ref{points:lacunae})
exceeds $2^n / (card(A) \cdot R) = 2^{t(m, A)}$. If such value $j_0$ exists, in view of bounds (\ref{points:lacunae-bounds}), $$\Sigma_j \ge 2^{head(m_{j_0})} - 2^{tail(m_{j_0})} \ge 2^{tail(m_{j_0})},$$ so lacunae would be large enough (to contain $1/R$ points on average) if $tail(m_{j_0})\ge t(m, A).$ Thus $tail(m_{j_0})$ serve as an alternative definition of the threshold. If there is no such $j_0,$ the threshold is set to $n$.

Assuming that the mask has components below the threshold, consider competition between the grasshopper and the frog. The difference is in the behavior below the threshold where lacunae correspond to lower contiguous components of the mask, and it amounts to comparing $(\overline{N_1}-\overline{N_2}) / R$ and $\overline{N_3}$. Here $\overline{N_2}$ is the number of jumps above the threshold. We can drop mask components above the threshold without affecting competition outcome and assume that $N_3=N_0$ and $N_2=0.$  So essentially we are again comparing frog and crawler behavior, however this time the condition ensuring frog's victory is not satisfied and the crawler (and hence grasshopper) has advantage. 

Overall, situation can become bad for the frog when the following takes place: the tail of the mask is low so that the sparse case kicks in; within lacunae there is at most one non-empty jump region. Then the frog would be forced to jump a number of times comparable to $card(A),$ which loses to sequential scan unless $R=1.$ For the grasshopper, however, this is not critical if there are higher mask components corresponding to the dense case. 

Note. As we have seen, if the matcher returns a negative mismatch value $-y$ for some $x$, it means that $x$ belongs to a lacuna within some fundamental region $T^y$ and the next
cluster of the PSP locus is located in the same fundamental region. If the mismatch is positive, it means that $x$ is in a larger lacuna located between two fundamental regions of 
order higher than $y$. This indicates that, in principle, the grasshopper could have operated two different thresholds and jumped more often upon encountering positive mismatch. 

Another possible variation of the algorithm is to try to enhance the scanning portion of it, by determining, upon seeing an element that qualifies, the end point of the cluster, in the PSP 
locus, to which it belongs, and then blindly picking the elements encountered before that end point. This means that instead of verifying the match, we will be verifying the inequality. 
These two operations have roughly the same cost. Besides, upon encountering an element that does not satisfy the inequality, we would still have to check if it matches the pattern. 
Calculating the end of the cluster is easy, but also bears additional cost. So upon first glance, this variation does not bring any benefit. However, its efficiency really depends on how
the appropriate storage interface is implemented. In many data stores, extra comparisons are made anyway. But some of the data store's costs may be avoided, 
if it is partitioned - the search simply stops at the end of the partition, and we may not even have to check that $x \le b$ in the loop.

\subsection{Partitioned case} \label{sec:partitioned}

In this section we describe grasshopper strategy modifications for Problem 2 (partitioned case).

If the data is partitioned, and it is possible to scan partitions in parallel, there is an obvious benefit to all of the strategies, because a bunch of crawlers, frogs and grasshoppers can
participate in the race and fill their bags faster. However, grasshoppers have the additional benefit of coming up with a proper threshold specific to a particular part.

When a partition is factorizable, there is a common pattern that all elements possess. The case of partitioning by intervals is discussed here as it is used most often. If an interval $L$ 
in $S$ is factorizable, there is a common prefix pattern $P$ and a corresponding prefix mask $M_L$ projecting on $d_L$ dimensions, such that $L = P | L'$, where $L'$ 
is an interval in an $(n-d_L)$-dimensional space. Prefix compression techniques are used by some stores to keep a single copy of the prefix and only $(n-d_L)$ bits per key. If the store
can also provide access to truncated keys (dimensionality reduction), efficiency increases. Unfortunately, unless the store is in our control, such access is often unavailable. 
As a result, the store performs multiple memory allocations and copies in order to assemble full-length keys which is counterproductive.

Nevertheless, computing the prefix from the boundaries of $L$ is easy, and additional reductions are possible. First, form $S' = S(m)\cap S(M_L)$. This is achieved through an easy mask
operation. If $S'\neq \emptyset$, let $m'$ be the corresponding intersection mask. If $p_{m'} \neq P_{m'}$, the entire interval $L$ lies outside the PSP locus (trivial mismatch), and can 
be safely skipped. If $m' = m$, it means that $S(M_L) \subset S(m)$, and hence the entire interval $L$ lies within the PSP locus (trivial match), so all the points in it are added to
the bag without checking. Otherwise, the problem is non-trivial, but the mask in PSP can be replaced with $m''=m \setminus m'$ and the pattern - with $p_{m''}$. When computing the threshold,
dimensionality, $n$, can be reduced by the dimensionality of $S(M_L)$. 

\subsection{Range queries}\label{sec:range}

In this section we describe geometric properties and matcher implementation for range restrictions. 

Before presenting appropriate geometric considerations for range queries, it makes sense to mention our reduction techniques.

We deal with pattern restrictions of kind (R): $x \& m \in [a,b].$ We first perform a trivial check $a\neq b$, otherwise it is a point restriction. Next we determine
if the interval is factorizable. For that we compute the maximal common prefix $p$ of $a$ and $b$. If such common prefix exists, then $[a,b] = [p | a', p | b'] = p | [a', b']$, and
all points within the interval have the same prefix $p$. This induces splitting of mask $m$ into prefix and suffix masks: $m = m_{prefix} | m_{suffix}$, and the original 
PSP is transformed into a system of two PSPs $x \& m_{prefix} = p$ and $x\& m_{suffix} \in [a',b']$. For the first problem we already know the locus structure, and locus of the 
original PSP is a subset of it. Hence considerations of the previous section would apply. For range specific techniques, it is then sufficient to consider the case of 
non-factorizable interval $[a,b]$ which is what we further assume. 

Observe that, by our assumption, elements $a$ and $b$ have different senior bits, 0 and 1 respectively (otherwise they have common prefix). 
An interval is called {\it complete}, if all bits of $a$ are 0 and all bits of $b$ are 1. Obviously, for a complete interval, the PSP is trivial: all elements of $A$ are solutions. 

An interval is called {\it suffix-complete} if it is factorizable, and its suffix interval $[a', b']$ is complete. For example, interval $[12, 15]$ is suffix-complete, since 
$[12, 15] = 12 | [0,3]$, but interval $[11,14]$ is not, since $[11, 14] = 8 | [3, 6]$. For suffix-complete intervals, via the mentioned reduction, the original range PSP is thereby 
converted into a point PSP. 

Assume finally that the interval is incomplete and non-factorizable. It may still sweep almost the entire corresponding $d$-dimensional subspace, and hence PSP locus may be
almost the entire space $S$. The smaller the interval, the more close the problem is to the point case, and the more chances for grasshopper strategy to find large lacunae to jump over.

Let us examine the corresponding geometry. As we have seen, the locus of the point PSP consists of intervals of equal length with gaps between them. This is not the case 
for range restrictions.

\begin{proposition}\label{prop:ranges}
Let $m$ be an arbitrary mask projecting onto $d$ dimensions and let $\{m_i\}$ be its canonical partition. Let $x\& m = [a,b]$ be a range PSP, 
$r$ be the cardinality of $[a,b]$, and let $r_i$ be the cardinality of $[a_{m_i},b_{m_i}]$. 

Then the locus of the PSP generally consists of clusters of varying lengths, which are separated by lacunae of total length
$spread (m, PSP) - r\cdot 2^{n-d}$. The spread can be calculated as $b | 1_{\ttilde m} - a | 0_{\ttilde m} + 1$. 
Individual lacunae lengths are partial sums 
\begin{equation}\label{ranges:lacunae} 
\Sigma_j = \sum_{i\ge j} [2^{head (m_i)} - r_i \cdot 2^{tail (m_i)}].
\end {equation}
\end{proposition}

Since the proof is more complex than in the point PSP case, we explain it a little bit.

First, suppose that mask $m$ is contiguous. In that case, as in the point case, within each fundamental region $T^{head(m)}$ in $S$, the locus of the PSP is a 
single interval of size $r\cdot 2^{tail(m)}$, where $r = b - a+1$ is the length of the interval. The lacuna between the intervals is thus $2^{head(m)} - r\cdot 2^{tail(m)}$. 

\begin{figure}
\centering
\includegraphics[width=3.2in,height=0.8in]{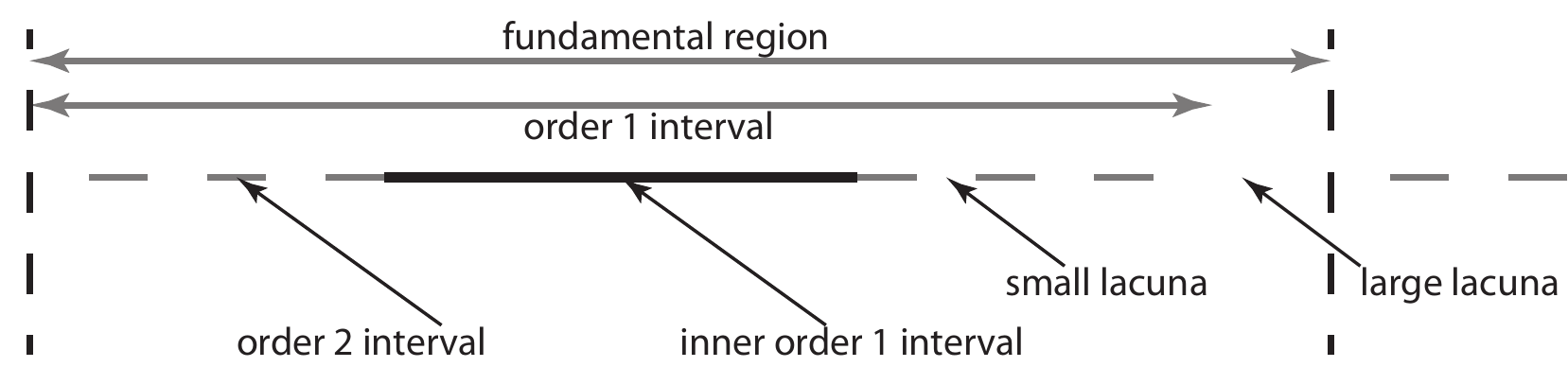}
\caption{Structure of the locus for range PSP.}
\label{fig:ranges}
\end{figure}

The picture becomes more complex in non-contiguous case. For simplicity, consider the case of two components.
Unlike the point case, the partial PSPs for each of the masks are not independent; the second PSP depends on the state of the first problem. 
Consider the PSP1: $x \& m_1 \in [a_{m_1}, b_{m_1}]$. Note that if $x \& m_1 \in (a_{m_1}, b_{m_1})$, then $x$ definitely solves the 
original PSP, and PSP for the second mask is not even considered. If $x \& m_1 < a_{m_1}$, or $x \& m_1 > b_{m_1}$, $x$ is definitely not a solution. If 
$x \& m_1 = a_{m_1}$ ($b_{m_1}$ respectively), 
then $x$ is a solution whenever it solves the second PSP (PSP2) of the form $x \& m_2 \in [a_{m_2}, 1_{m_2}]$ ($x \& m_2 \in [0_{m_2}, b_{m_2}]$ respectively). 
Denote these PSPs as $PSP2 (a)$ and $PSP2 (b)$ respectively. Of course, one or both of the corresponding intervals may degenerate to a point.

%Note that if the first component $m_1$ has a single bit, PSP1 never determines whether $x$ solves the problem since, as we observed, the senior bits of $a$ and $b$ are always different. 
Let $r_2(a)$ and $r_2(b)$ be the lengths of the intervals involved in $PSP2(a)$ and $PSP2(b)$. As we have mentioned, within each 
fundamental region $T^{head(m_1)}$, the locus of PSP1 consists of a single interval of length $r_1\cdot 2^{tail(m_1)}$, called {\it order 1} interval. The locus of the original 
PSP within that fundamental region is contained in that interval, and contains the interval of length $(r_1-2)\cdot 2^{tail(m_1)}$ which corresponds to the inner part 
$(a_{m_1}, b_{m_1}) = [a_{m_1}+1, b_{m_1}-1]$, if the latter is not empty. Within the interval of order 1 but outside this inner part, in every fundamental region 
$T^{head(m_2)}$, there are two series of {\it order 2} intervals corresponding to PSP2(a) and PSP2(b), located to the left and to the right respectively 
of the inner part of the order 1 interval. One of the order 2 intervals in each of two series would be adjacent to the order 1 interval from the corresponding side, 
and the total number of intervals within that fundamental domain would be at most $2 \cdot (2^{tail(m_1)- head(m_2)} - 1)$. See figure \ref{fig:ranges} for illustration. 

The order 2 intervals in each series have lengths $r_2(a)\cdot 2^{tail(m_2)}$ and $r_2(b)\cdot 2^{tail(m_2)}$ respectively, and lacunae between these intervals are of sizes 
$2^{head(m_2)} - r_2(a)\cdot 2^{tail(m_2)}$ and
$2^{head(m_2)} - r_2(b)\cdot 2^{tail(m_2)}$ respectively. Now, there was a lacuna between the order 1 intervals in different fundamental regions $T^{head(m_1)}$. 
To this gap one needs to add the gap between the beginning of that interval and the closest intervals of the order2 interval series from each such fundamental region. 
This gap corresponds to intervals $[0_{m_2},  a_2-1]$ and $[b_2+1, 1_{m_2}]$, and has combined length 
$(a_2 + 2^{d_2} - b_2 -1) \cdot 2^{tail(m_2)} = 2^{head(m_2)} - r_2\cdot 2^{tail(m_2)}$, where $r_2 = b_2-a_2+1$ is the length of $[a_2, b_2]$. Adding the gap to the 
length of the order 1 lacuna produces the partial sum (\ref{ranges:lacunae}).

If the mask has 3 components, the picture changes in a similar manner: each of the order 2 intervals would have its inner part belonging to the PSP locus,
and within the space between order 2 interval and its inner part there would be two series of order 3 intervals, and so on.

We now describe how the matcher works for the range PSP case. 

For the {\tt Mismatch} operation, the matcher examines each $PSP(m_i, [a_i, b_i])$, $i=1,\ldots$, one at a time. If $x\& m_i \in (a_i, b_i)$, the matcher returns 0, 
indicating a match. If $x\& m_i \not\in [a_i, b_i]$, let $e_j$ be the most senior bit, 
on which they disagree. The matcher returns $j$ if $x\& m_i > b_i$ and $-j$ if $x\& m_i < a_i$. If $x\& m_i = a_i$ or $x\& m_i = b_i$, the matcher proceeds on to 
$PSP(m_{i+1}, [a_{i+1}, b_{i+1}])$, where the interval is either $[a_{m_{i+1}}, 1_{m_{i+1}}]$ or $[0_{m_{i+1}}, b_{m_{i+1}}]$, and so on. 

Let $I$ be the identity mask on $S$, i.e. the mask, projecting onto entire $S$.

For the {\tt Hint} operation, given an element $x\in S$ and mismatch position $j$, the matcher acts as follows. 

If mismatch is negative, the matcher computes a preliminary hint $h_1$ of the form $x_{I_{>j}} | 1_{I_{=j}} | 0_{m_{<j}} | 0_{\ttilde m_{<j}}$. If preliminary hint is not 
within $[a,b]$ (which depends on the senior part of $x$), the hint is corrected as $h_1 | a_{m_{<j}}$. The highest bit position that changes is $j$. 

If the mismatch is positive, the matcher determines the ``growth point" $g$ which is the smallest position, above $j$, of an unset (0) 
bit in $x\& (\ttilde m)_{>j}$. If such position does not exist, the search is over ($\infty$ returned), otherwise the hint is computed as above with $g$ instead of $j$. 

The way to compute the threshold for the grasshopper strategy is similar to the point queries case, but uses Proposition \ref{prop:ranges}. The threshold would be $tail(m_{j_0})$ 
where $j_0$ is the maximal $j$ for which $\Sigma_j$ in (\ref{ranges:lacunae}) exceeds $2^n / (card(A) \cdot R).$

Treatment of the partitioned case is similar to point PSPs, but for each interval in the partition we can also compute a proper range restriction. 
For a given interval, besides the trivial cases when the entire interval qualifies, or the entire interval does not qualify, there may be other interesting situations, 
when, e.g. range restriction becomes point restriction within the interval, or when restriction with incomplete range becomes a restriction with complete range, and so on.

\subsection{Set queries}\label{sec:set}

In this section we describe geometric properties and matcher implementation for set restrictions. 
We have similar reduction techniques for them. 

We now deal with pattern restrictions of the kind (S): $x \& m \in E,$ where $E$ is some set. For convenience, we assume the set to be ordered. We first check that the spread of
$E$ is not equal to its cardinality, otherwise $E$ is actually a range. This also excludes single element sets. Next we determine if the set is factorizable. For that we compute the
maximal common pattern $p$ of all elements of $E$. If such common pattern exists, then $E = p | E'$, with the splitting of the mask $m = m_{common} | m_{residue}$. The original PSP is 
transformed into a system of two PSPs $x \& m_{common} = p$ and $x\& m_{residue} \in E'$. For the first problem we already know the locus structure, and locus of the 
original PSP is a subset of it. Hence considerations of the point queries section would apply. For set specific techniques, it is then sufficient to consider the case of 
non-factorizable set $E$, which is what we further assume. 

Since a set consists of individual points, it is clear that the locus of the set PSP is a union of loci of corresponding point PSPs. This suggests that all clusters in PSP
locus have the same size, and their total number differs from the point case by a factor $card(E)$. As in the range case, 
the solution space can be quite large if the set is almost the entire space $S(m)$. Moreover, individual lacunae sizes differ greatly depending on the distances between 
the set elements. However since the set is fully contained in the range $[min(E), max(E)]$, estimates of the lacunae around the edges of the appropriate fundamental regions
are similar to the range case. If they are not large enough to justify hopping, we still have the option to look for sufficiently large lacunae corresponding to gaps between
set elements.

The matcher does not split the locus of the set PSP into the union of point PSP. Instead, it splits the set PSP into similar partial set PSPs by components of the mask partition.

As in the range case, each next PSP depends on the state of the previous one. For the first PSP with restriction $x \& m_1 \in E_1$, 
if $x\& m_1 \not\in E_1$, search is immediately interrupted as a clear mismatch. If $y=x\& m_1 \in E_1$, further search is reduced to the subset $E_2(y)$ of $E$ that matches $y$ 
as the prefix (the next PSP would then be $x\& m_2 = E_2$ with $E_2 = E_2 (y)_{m_2}$). The matcher keeps track of all such elements and perhaps the one immediately below them 
in order to determine
the correct mismatch position, from above or from below. With each next PSP, the cardinality of $E_i$ quickly reduces. 

Similarly, when providing hints, the matcher must find the appropriate smallest element to which it can move from the current position.

Similar to range PSPs, partitioning by intervals brings new aspects as, for a particular interval, the set PSP may morph into a range or point PSP, and so on. 

\subsection{Multiple pattern search}\label{sec:combine}

We have already developed matchers for each kind of filters. Here we outline, how multiple simultaneous restrictions are handled. 

Since the locus of simultaneous PSP is the intersection of the loci of the individual PSPs, the lengths of the lacunae add up. So it is possible to set a single 
threshold, but calculation of it is more involved. 

When there are multiple pattern restrictions of various kinds, the matcher starts with performing reductions described in the previous sections. All resulting fixed patterns,
from point filters and from factorization of range and set queries, are combined into a single fixed pattern. All complete residual interval PSPs are 
eliminated, etc. The matcher is left with possibly a single point PSP and/or possibly multiple range and set PSPs. The matcher then employs individual matchers
for each PSP and makes them compete for the highest mismatch position. When mismatch is given, the hint is computed to satisfy all restrictions at the same time. 
If the mismatch is negative, a preliminary hint is computed, and each individual matcher corrects it if necessary. When mismatch is positive, all matchers compete 
for the lowest growth position, and then proceed as for negative mismatch.  

\section{Experimental Results}\label{sec:results}

We decided to test the Grasshopper strategy with various thresholds against the crawler strategy, since the purpose was to gain advantage over full scans. We tried in particular threshold 0 (frog strategy). With lower threshold the number of hops increases at the expense of shorter jumps.

\subsection{Implementation}\label{sec:implementation}

The matcher code was written in Java with keys represented as byte arrays. We had to implement unsigned large integer arithmetics and bitwise operations on them. The matcher code was rewritten a few times to achieve very low cost of its operations. API were used to create the schema and the query filters. 
For the data store, we created a pluggable storage adapter interface to experiment with different stores. 

We tested the distributed data scenario (on Hadoop) and the in-memory scenario. The former was important to confirm that Grasshopper algorithms could be helpful in Big Data analytics; the latter would be beneficial for in-memory databases including embedded ones. Both scenarios have been receiving much publicity these days.

We created data store adapters for in-memory scenario based on standard Java {\tt TreeMap} and a simple B+-tree that we wrote ourselves. 

For the big data tests, we picked Apache HBase \cite{opensource:hbase}, an open source distributed key-value store from the Hadoop family. 
Within this adapter, Grasshopper algorithms were invoked via the HBase coprocessor mechanism. HBase partitions data into key ranges (regions), each of them assigned to a region server node. Coprocessors allow access to each region which in turn enables partition-based grasshopper strategies. We also had to implement another coprocessor that kept track of statistics for every region.

\subsection{Hardware and schema setup}\label{sec:hardware}
For in-memory testing, we used a standard Thinkpad laptop with an i5 CPU and 16 Gb of RAM running 
64 bit Windows 7 OS. The data was either randomly generated on the fly or read from a file. The schema emulated call detail records (CDR) typically produced by a telecom. 
There were 16 dimensional attributes of various sizes ranging from 2 to $2^{14}$. The total composite key length in that case was 116, resulting in 15 byte keys. 
A data set of 100 mln records was used, since it was hard to fit larger volumes into memory. The maximal Java heap size was set to 12 Gb. All 
in-memory tests described below were run single-threaded, except a few using partitioned B+-tree. 

For distributed storage tests, we used a configuration with 128 regions on 12 region server nodes running version 0.94 of HBase on Hadoop installed on a commodity Linux cluster. We tried several data sets: one with the mentioned CDR schema but with 150 mln records, one with 10 attributes and with 1.46 bln records, also related to telecom, and a TPC-DS 
benchmark data set with 5 attributes and 550 mln records. Our cluster, unfortunately, did not have enough disk storage to host larger data sets.

The queries could be expressed in SQL
as {\tt SELECT COUNT(1) FROM dataset WHERE filter}, with filter being a point, range or set restriction on some of the dimensional attributes of the dataset. 
Such queries suit our purposes best, and they are generally useful, e.g., in data mining scenarios.

\subsection{How we tested}\label{sec:how}

For all internal tests, query filter values were randomly generated on the fly. For in memory scenario, we ran exhaustive combinations of queries for up to 3 attribute filters with point, range and set restrictions. For big data scenario, attributes were always chosen randomly. Each query was run 10 times, using crawler and grasshopper strategy with different thresholds (including threshold 0, or, frog strategy), then the smallest and the largest run times were eliminated and an average of the remaining runs was computed for each strategy. After that we computed 
the average over all combinations and compared the results.

Besides varying thresholds, we also tried different strategies for composing the key. 

\subsection{Results and discussion}\label{sec:discussion}

Odometer key composition strategies for leading 
attributes produced, as expected, very good performance, but for far too many other cases the grasshopper had to resort to pure crawling. It was still efficient overall; 
however we determined that for 
ad-hoc queries a much better choice was to do single bit interleaving in the decreasing order of attribute cardinalities. In most cases we could then speed up ad-hoc queries 
on {\it every} attribute. As said, our methods provide improvements over full scan for any gz-curve composition kind, but not necessarily for each attribute. 
If improvement is desirable for more attributes, we strongly recommend using single bit interleaving for key composition. 

As dimensionality grows, 
the number of attributes that can take advantage of grasshopper techniques will be limited (to those, whose mask heads are high enough; with single bit interleaving more 
attributes would fall into that category). According to Proposition \ref{prop:grasshopper}, the number $w$ of ``useful" bits in the key is roughly $\log_2(card(A)\cdot R)$, so setting the threshold at $n-w$ is close to the theoretically best option. Those $w$ bits must be distributed between the most popular attributes. 
Optimal key composition is not discussed in this paper. All our results below are shown for single bit interleaving. 

In our experiments, with the exception of one data set, the best results were achieved for positive thresholds, i.e. the grasshopper was faster than the frog. For in memory data sets, the frog was on average 3-5 times slower than the crawler while the grasshopper was faster than the crawler. For distributed data sets, both the frog and the grasshopper outperformed the crawler by orders of magnitude. With optimal threshold, the grasshopper was 6.5\% faster than the frog on the CDR data set, 13\% faster on the TPC-DS data set and for the 1.45bln records data set both strategies coincided (threshold 0). We decided to exclude frog's timings from the charts for better visualization.

\begin{figure}
\centering
\includegraphics[width=3.2in,height=1.4in]{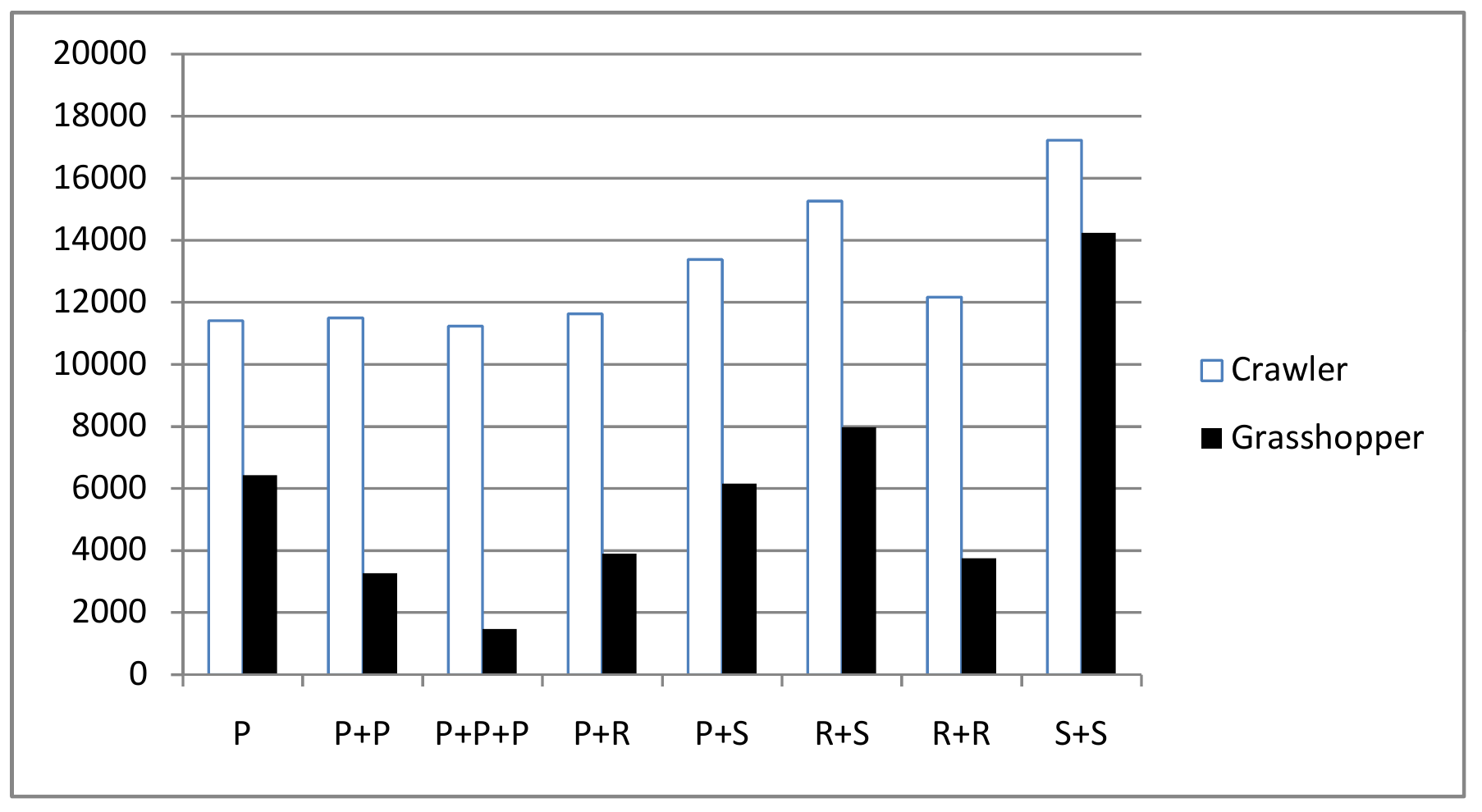}
\caption{Query times (ms) with TreeMap as data store. Combinations of filters include point (P), range (R) and set (S) restrictions
16 dimension, 100 mln rows data set. Measured using exhaustive combinations. The frog (not shown) is on average 4.3 times slower than the crawler.}
\label{fig:treemap}
\end{figure}

\begin{figure}
\centering
\includegraphics[width=3.2in,height=1.4in]{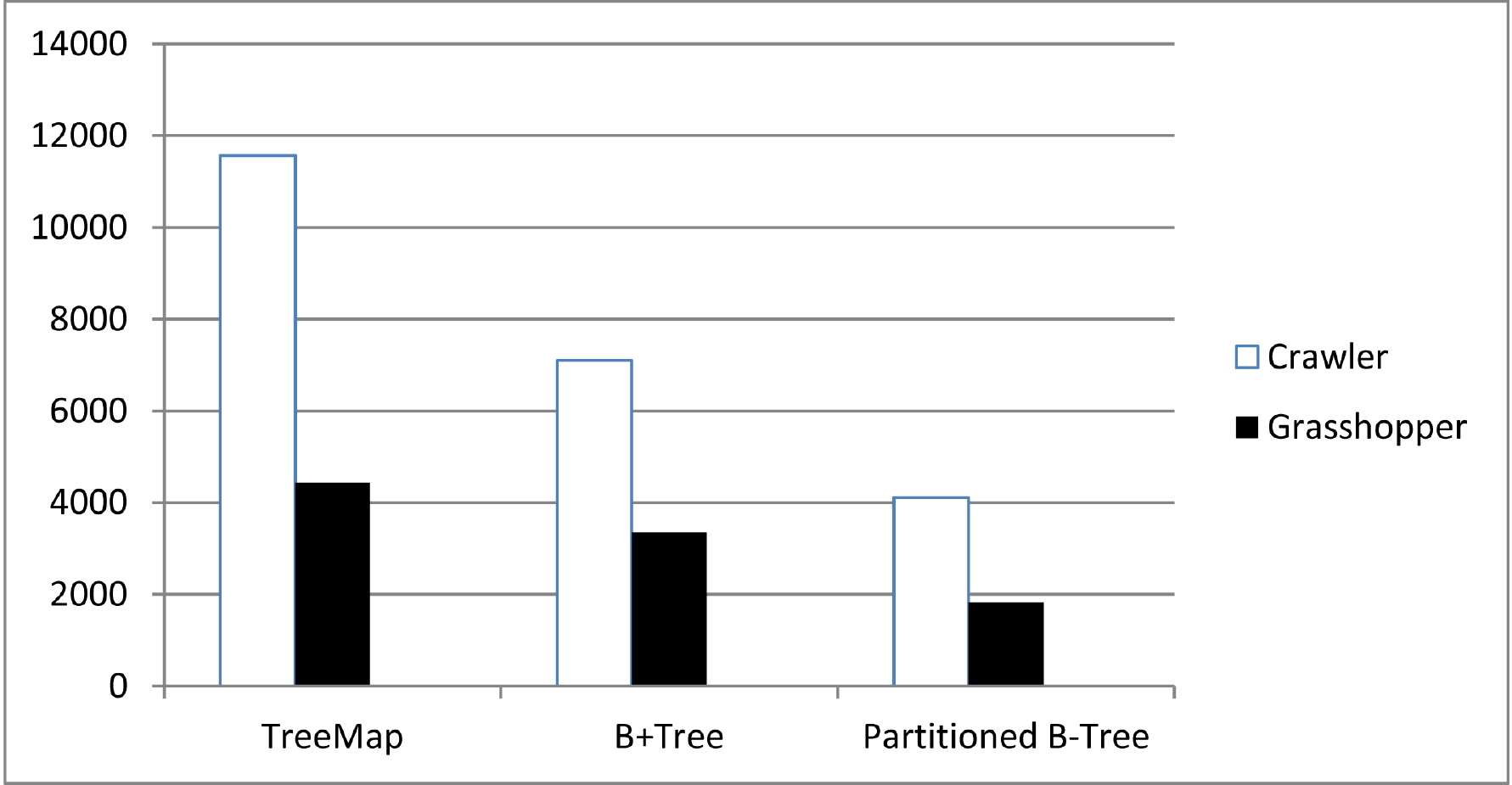}
\caption{Query times (ms) with TreeMap, basic B+-tree and partitioned B+-tree as data store for single point filters on 16 dimension, 100 mln rows data set. 
Measured using exhaustive combinations. The frog (not shown) is at least 3.8 times slower than the crawler.}
\label{fig:tmmvbt}
\end{figure}

The grasshopper with appropriately chosen threshold never lost to the crawler on any of the tested data sets. 

The TreeMap was the fastest in data load, but lost to some B-tree based implementations in query performance and consumed more memory. 
Nevertheless we feel that for custom-programmed in-memory OLAP scenarios it is a very reasonable choice. 
Results for different filter kinds are shown in figure \ref{fig:treemap}. The more restrictions there are, the larger the performance gains of Grasshopper strategy. 

Comparison of query times on the same data between in-memory data stores is shown in figure \ref{fig:tmmvbt}; all data stores greatly benefited from using Grasshopper strategy. 

For all in-memory tests, we found the theoretically computed threshold for grasshopper jumps to be the best in the majority of cases. We measured the scan-to-seek ratio $R$ using various techniques, and found it to be ranging from 0.35 to 0.8 for in-memory stores. For the 100mln CDR data set, for example, the theoretical threshold was close to 95. Thus 21 (116-95) key bits were ``useful", and all 16 dimensions could benefit from Grasshopper strategy.

It was hard to compute the scan-to-seek ratio for distributed case because it seemed to vary quite significantly. One of the techniques we used did produce optimal threshold values for two data sets, but certainly better methodology is needed. The optimal threshold value for the 150mln CDR data set in this case was 64 (or, 52 ``useful" bits). In many cases it was enough to assume $R=1.$

\begin{figure}
\centering
\includegraphics[width=3.2in,height=1.4in]{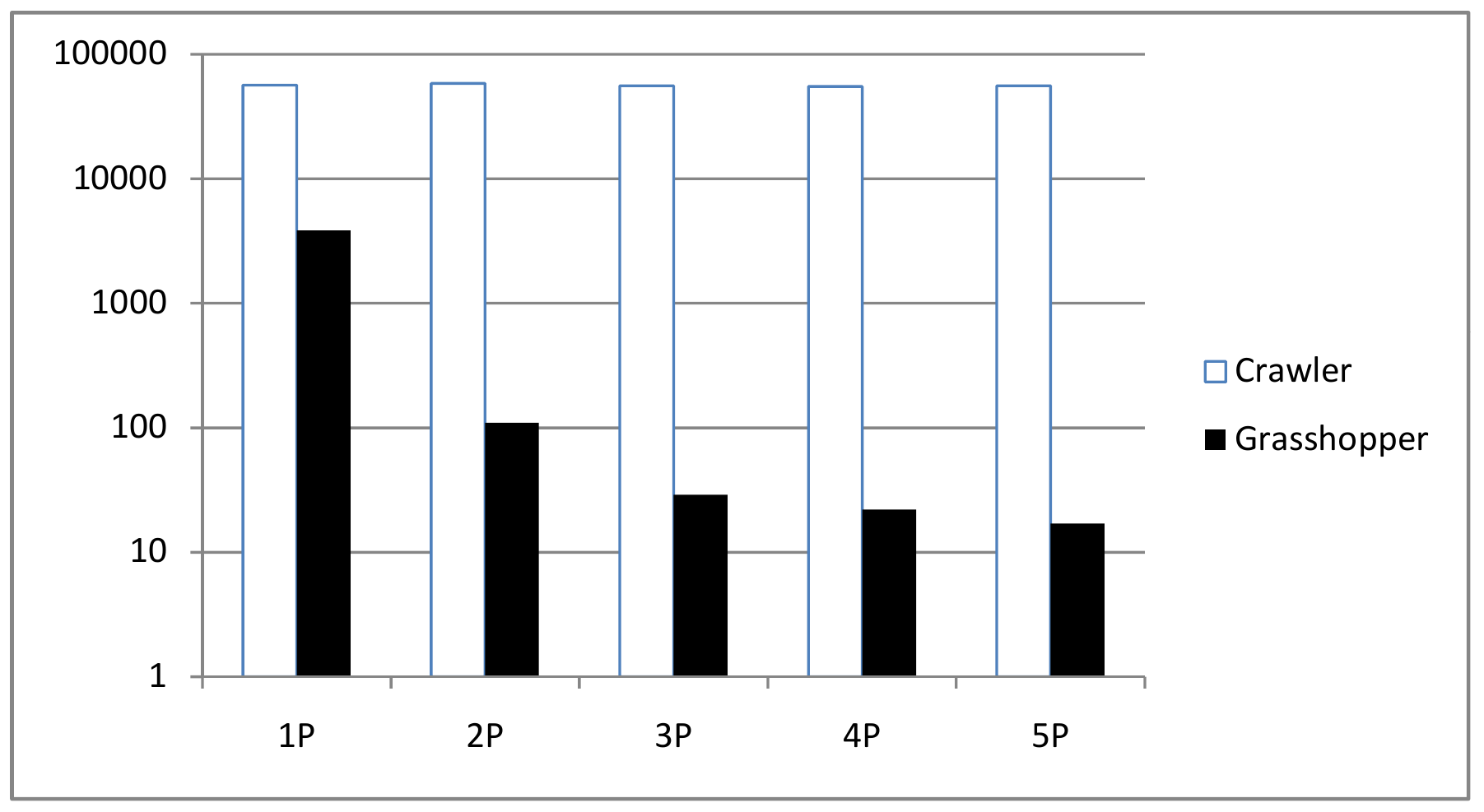}
\caption{Query time (ms) comparison on logarithmic scale using HBase as data store for single and 
multiple point filters on 10 dimension, 1.46 bln rows data set. Measured using random combinations. Here threshold=0, i.e. grasshopper and frog strategies coincide.}
\label{fig:sdr}
\end{figure}

\begin{figure}
\centering
\includegraphics[width=3.2in,height=1.4in]{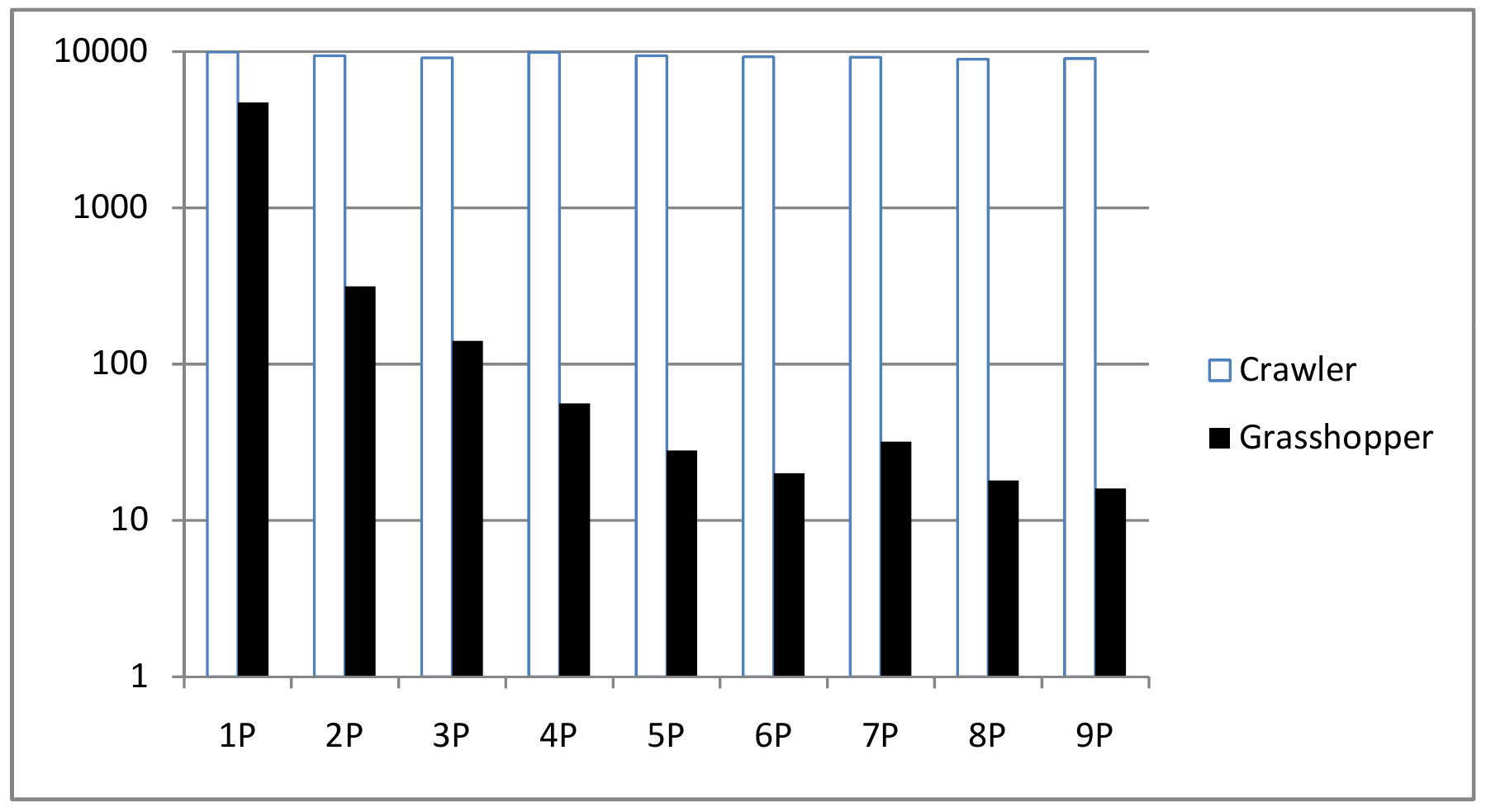}
\caption{Query time (ms) comparison on logarithmic scale using HBase as data store for single and 
multiple point filters on 5 dimension, 550 mln rows TPC-DS data set. Measured using random combinations. The frog (not shown) is on average 13\% slower than the grasshopper.}
\label{fig:tpc}
\end{figure}

\begin{figure*}
\centering
\includegraphics[width=5in,height=2.9in]{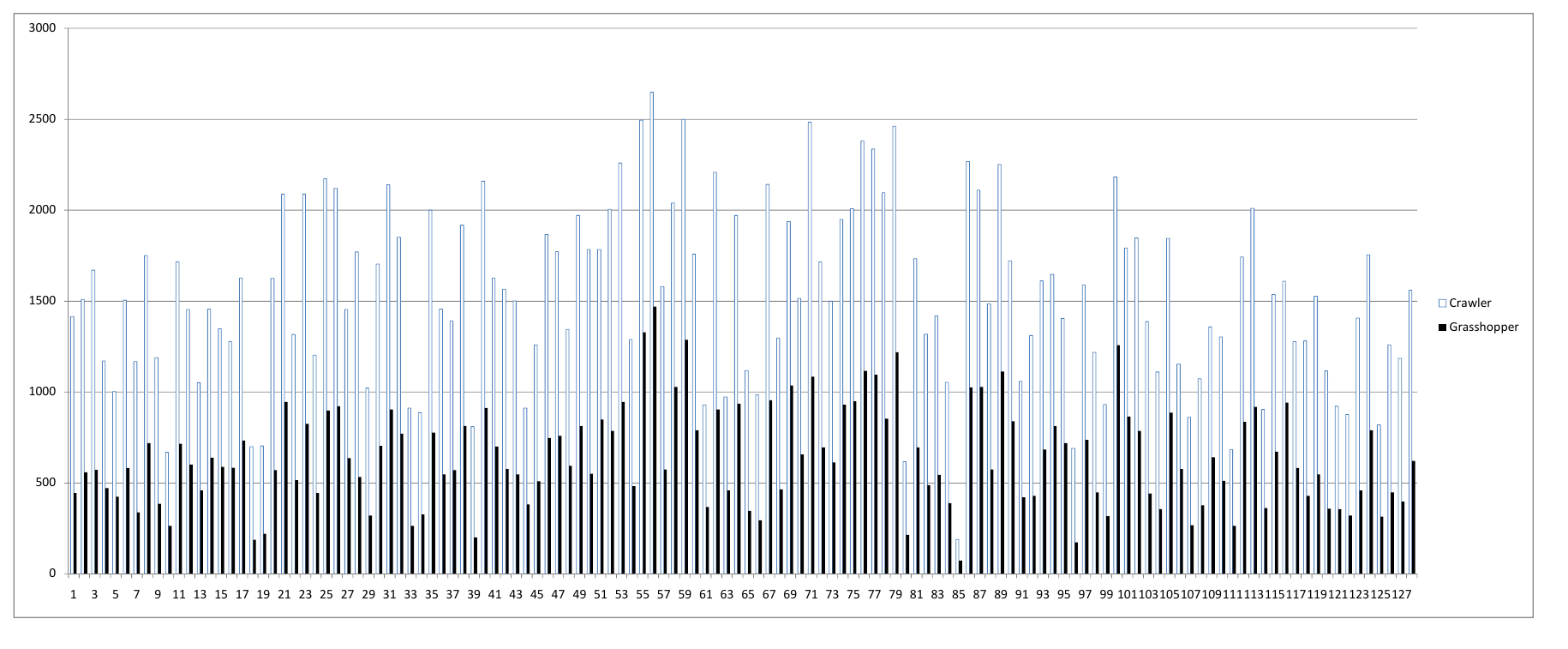}
\caption{Query times (ms) per HBase region with HBase as data store for single point filters on 16 dimension, 150 mln rows data set. Measured using random combinations. The frog (not shown) is on average 6.5\% slower than the grasshopper.}
\label{fig:cdr}
\end{figure*}

A few comments are due for big data tests. 
In HBase, region data is split internally into blocks. Skipping over a block is beneficial, but block statistics are not accessible from the coprocessor. Searches within the block in version 0.94 are 
sequential, so seek operation is very slow unless it skips over entire block(s). This improved in the version 0.96 of HBase, to grasshopper's advantage. Although nodes had enough memory,
despite using recommended settings, it looked unlikely (and impossible to verify) that all data was cached. Nevertheless on some tests grasshopper was so fast, that we had to use 
logarithmic scale in time comparisons.

The time of query completion is determined by the slowest node, and if data is not evenly distributed, results are less predictable. 
Test time differences of both strategies per region on the CDR data set  are presented in figure \ref{fig:cdr}. The results for the TPC-DS data set are presented in figure  \ref{fig:tpc}; figure \ref{fig:sdr} shows results for the 1.45bln record set. 

In all cases, Grasshopper algorithms helped quite a bit, often by many orders of magnitude. 

\subsection{Ad-hoc query competition}\label{sec:benchmark}

As our prototype matured, we were given an opportunity to participate in an ad-hoc query benchmark competition handled externally by a different department in our HQ. Participants included also a top commercial distributed RDBMS and a distributed MPP database built on top of PostgreSQL (both written in C/C++ and not Hadoop-based). The test was against a 12 bln record CDR data set, and the participants were allowed to build any indices they wanted, however queries were not disclosed. The cluster had 3 high-end servers, each with 204 Gb of RAM and two 6-core CPUs running Linux. The 16-query test had two metrics: total query time and maximal query time. Each query included at least one filter on dimensions.

By competition time our prototype system had an SQL front end, a B-tree based distributed cache for keys and a version of HBase 0.96 with the following changes: data storage  split over two formats - keys in HBase "hfile" format, values in columnar "rcfile" format \cite{he:rcfile}, and data access directly via HDFS readers. We did not build any indices.

For the competition test, choice of queries was out of our control. Organizers took best times over 10 runs of all queries.

\begin{figure}
\centering
\includegraphics[width=3.2in,height=0.4in]{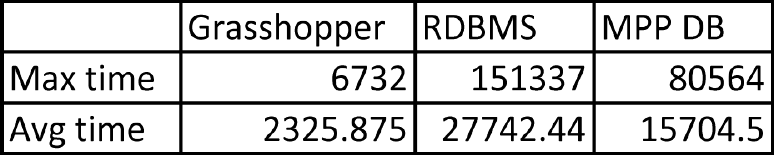}
\caption{Query time (ms) comparison of Grasshopper using tweaked HBase as data store vs commercial RDBMS and PostgreSQL-based MPP DB with random
point and range filters on 18 dimension, 12 bln rows CDR data set. }
\label{fig:benchmark}
\end{figure}

The benchmark results (figure \ref{fig:benchmark}) showed huge advantage of grasshopper methods over commercial RDBMS and PostgreSQL-based MPP DB. 

\section{Conclusions and Future Work}

The paper presents Grasshopper algorithms which allow significant improvement in performance of ad-hoc OLAP queries with point, range and set filters, without any additional indices 
or materialized views. The algorithms are applied after a reduction of storage to key-value form using composite keys with generalized z-curves encoding. Their main field of 
application is Big Data, for ad-hoc analysis of large in-memory or distributed data sets. 

There are several directions which are interesting to investigate in this regard: cooperative scanning, floating length keys, including binning attributes into the key, 
applications to data mining, streaming versions, etc.

\section{Acknowledgments}
The author would like to thank Sergey Golovko and Vladimir Rozov for numerous discussions, 
helping to bring the grasshopper alive and to test its abilities, and Yan Zhou for 
carefully proofreading preliminary versions of the paper.

\bibliographystyle{abbrv}
\bibliography{arxiv_grasshopper}

\end{document}